\documentclass[prl,twocolumn,superscriptaddress]{revtex4}
\usepackage{amsmath,amssymb,mathrsfs}
\usepackage[dvips]{graphicx}
\usepackage{hyperref}
\usepackage{paralist}

\newcommand{\ua}{\uparrow}
\newcommand{\da}{\downarrow}

\newcommand{\Li}{\mathrm{Li}}
\def\be{\begin{equation}}
\def\ee{\end{equation}}
\def\bea{\begin{eqnarray}}
\def\eea{\end{eqnarray}}

\def\scr{\mathscr}
\def\SG{{\scr G}}

\begin{document}

\title{Analytic thermodynamics and thermometry of
Gaudin-Yang Fermi gases}

\author{Erhai Zhao}
\affiliation{Department of Physics and Astronomy, University of
Pittsburgh, Pittsburgh, Pennsylvania 15260, USA}

\author{Xi-Wen Guan}
\affiliation{Department of Theoretical Physics,
Research School of Physics and Engineering,
Australian National University, Canberra ACT 0200, Australia}
\affiliation{Mathematical Sciences Institute, Australian
National University, Canberra ACT 0200, Australia}

\author{W. Vincent Liu}
\email[e-mail:]{w.vincent.liu@gmail.com}
\affiliation{Department of Physics and Astronomy, University of
Pittsburgh, Pittsburgh, Pennsylvania 15260, USA}

\author{M. T. Batchelor}
\affiliation{Department of Theoretical Physics,
Research School of Physics and Engineering,
Australian National University, Canberra ACT 0200, Australia}
\affiliation{Mathematical Sciences Institute, Australian
National University, Canberra ACT 0200, Australia}

\author{Masaki Oshikawa}
\affiliation{Institute for Solid State Physics, University of Tokyo,
  Kashiwa 277-8581 Japan}

\date{\today}

\begin{abstract}
We study the thermodynamics of a one-dimensional attractive Fermi gas
(the Gaudin-Yang model) with spin imbalance. The exact solution has
been known from the thermodynamic Bethe ansatz for decades, but it
involves an infinite number of coupled nonlinear integral equations
whose physics is difficult to extract. Here the solution is analytically
reduced to a simple, powerful set of four algebraic equations.  The
simplified equations become universal and exact in the experimental
regime of strong interaction and relatively low temperature. Using the
new formulation, we discuss the qualitative features of
finite-temperature crossover and make quantitative predictions on the
density profiles in traps. We propose a practical two-stage scheme to
achieve accurate thermometry for a trapped spin-imbalanced Fermi gas.
\end{abstract}


\maketitle

Alkali fermionic atoms at tens of nano-Kelvin are a remarkable
addition to correlated quantum matter. Their interaction is
unprecedentedly tunable from attractive to repulsive infinity.  This
allows, for instance, the observation of universal properties of Fermi
superfluids at unitarity. Recent
developments~\cite{Zwierlein06,Partridge06} provide imbalanced
populations of cold atoms in different hyperfine (spin) states, adding
a new dimension in the low-temperature phase diagram~\cite{simons}.
In low dimensions, such attractive Fermi gases with spin imbalance
give rise to a series of very interesting phases. For example, it was
suggested that the quasi one-dimensional imbalanced Fermi gas, i.e., a
weakly coupled array of one-dimensional (1D) attractive Fermi gases,
gives a better chance to observe a long-sought crystalline superfluid,
known as the Fulde-Ferrell-Larkin-Ovchinnikov (FFLO) state
\cite{parish:07,zhao:08,KunYang:01}.

The realization of 1D ultracold atomic gases
\cite{PhysRevLett.94.210401,randy} provides a new setting to
experimentally test exactly solved models of interacting fermions.  In
particular, spin imbalance enables us to reach parameter regimes
impossible in solid state (such as quantum wires) to study exotic
quantum liquids outside the paradigm of a spin-charge separated
Tomonaga-Luttinger liquid (TLL).  Since their thermodynamics can be
computed exactly, 1D ultracold Fermi gases can also serve
as a calibration reference system to measure the thermodynamics of
strongly interacting Fermi gases in higher dimensions.

Contrary to the impression one might have, exact ``solvability''
(integrability) does not guarantee that physical quantities of
interest can be actually calculated.  
In principle, the thermodynamic Bethe
ansatz (TBA)~\cite{Takahashi} gives exact
results for thermodynamic properties.
However, it requires numerically solving an infinite number of coupled nonlinear integral 
equations to compare with experiments \cite{kakashvili:041603}.  
On the other hand, quantum Monte Carlo calculations are limited to small system sizes
 \cite{casula:033607}.

In this paper, we give a new formulation of the exact thermodynamics
of a 1D Fermi gas with contact attraction,
the Gaudin-Yang model \cite{Gaudin196755,PhysRevLett.19.1312}, which
is now accessible in cold atom experiments \cite{PhysRevLett.94.210401,randy}.
Based on the analytical analysis of the TBA equations,
we derive a simple, complete set of algebraic equations
which yield all thermodynamic quantities
and show that they are asymptotically exact and universal in the physically
interesting regime.
Our approach can be applied to a wide range of Bethe ansatz
integrable many-body systems.
As an application, we propose a two-stage scheme
to achieve accurate thermometry for trapped 1D Fermi gases.

{\sf Model and strong coupling limit.}
In the on-going experiments, for instance, at Rice University \cite{randy},
a system of parallel 1D gas ``tubes'' is prepared by
loading alkali fermionic atoms
(e.g., $^6$Li in hyperfine states $|F=1/2$, $m_F=\pm 1/2\rangle$) in a
square optical lattice.
The transverse dynamics is suppressed by deep lattice potentials, and
the motion is restricted along the tube axis $x$. The Fermi gas confined in
each tube is then well described by {the Gaudin-Yang model}
\cite{Gaudin196755,PhysRevLett.19.1312},
\bea
H=&-&\int dx \sum_{\sigma=\ua,\da} \psi^\dagger_\sigma(x)
\left[\frac{\hbar^2}{2m}\frac{\partial^2}{\partial x^2}
  +\mu_\sigma\right ]\psi_\sigma(x) \nonumber   \\
&-& g\int dx \psi^\dagger_\ua(x)\psi^\dagger_\da (x)\psi_\da
(x)\psi_\ua (x).
\eea
The two hyperfine species with equal mass $m$ are labeled with spin up and
down respectively. {\bf  $g>0$} is the contact attractive interaction.
In general, the chemical potentials for the {spin up and  down}
fermions are different, say $\mu_\ua>\mu_\da$.
Following convention, we define the chemical potential
$\mu=(\mu_\ua+\mu_\da)/2$, the effective magnetic field
$h=(\mu_\ua-\mu_\da)/2$, the total density $n=n_\ua+n_\da$, the
magnetization $M=n_\ua-n_\da$, and the polarization $P=M/n$. There
are two length scales in this problem: the 1D scattering length
$a_{1}=2\hbar^2/(mg)$ characterizing the interaction strength $g$, and
the inter-particle spacing $1/n$. Their ratio defines the
dimensionless interaction strength $\gamma=2/(n a_{1})$. In the
experiments, the gas is usually {dilute and strongly
interacting,} namely $a_{1}\ll 1/n$, so $\gamma\gg 1$. We
will focus
on this {\it strong coupling limit} where the binding
energy $\varepsilon_B=\hbar^2/(ma^2_{1})$ is much larger than the
chemical potentials.

The Gaudin-Yang model is exactly solvable by
means of the Bethe ansatz \cite{Gaudin196755,PhysRevLett.19.1312}.
Its zero temperature phase diagram has been worked out theoretically
\cite{orso:070402,hu:070403,guan:085120}. There are three
phases:  the fully paired (BCS) phase which is a quasi-condensate with
zero polarization, the fully polarized (normal) phase with $P=1$, and the
partially polarized (1D FFLO) phase where $0<P<1$ and the pair
correlation function oscillates in space \cite{zhao:08}.
For given $\mu$, the FFLO phase is separated from the BCS phase and
the normal phase by two quantum critical points at $h=h_{c1}$ and
$h_{c2}$, respectively.
An effective field theory of the 1D
FFLO phase, recently established by two of us, shows that it is a
novel type of two-component TLL,  each
gapless normal mode being a hybridization of spin and charge~\cite{zhao:08}.

Here, based on the exact Bethe ansatz solution, we demonstrate that,
in the low-energy limit the 1D FFLO is equivalent to two coupled gases
of {\it spinless} fermions.  Due to strong attraction, any minority
(spin $\da$) fermion would like to pair up with a majority (spin
$\ua$) fermion. Then, roughly speaking, the FFLO phase can be viewed
as a mixture of tightly bound pairs (labeled by subscript $b$) and
unpaired leftover fermions (labeled by subscript $u$)
\cite{orso:070402,hu:070403,guan:085120}.  By a transmutation in
statistics, the bosonic pair degree of freedom is equivalent to a gas
of spinless fermions, analogous to the well known case of
Tonks-Girardeau gas of hard-core bosons. The two gases are coupled by
residue scattering, so the effective chemical potential of each gas,
$\mu_b$ ($\mu_u$), depends on the Fermi pressure of the other gas,
$p_u$ ($p_b$).

{\sf Exact thermodynamics.}
Our analysis begins with recognizing the separation
of energy scales in the strong coupling limit.
At very low temperatures, $T\ll \mu_u,\mu_b$
(Boltzmann's constant $k_B=1$), the thermodynamics of the
1D FFLO phase is governed by the linearly dispersing phonon modes,
i.e., the long wavelength
density fluctuations of the two weakly coupled gases.
Using the effective field theory and the standard
conformal mapping~\cite{PhysRevLett.56.742,PhysRevLett.56.746},
we find the low temperature
thermodynamic potential (per unit length),
\be
\SG_{FT} (T)=\SG_0-\frac{\pi}{6\hbar}(\frac{1}{v_b}+\frac{1}{v_u})T^2 + ...\,.
\label{conformal}
\ee
The system has two gapless excitations, which are the
long wavelength
density fluctuations (phonons) of the two gases.
Their group velocities are the
respective Fermi velocities, $v_b = (1-P)[1+(1+3P)/\gamma]v_F/4$
and $v_u = P[1+4(1-P)/\gamma]v_F$, with
$v_F={\hbar n\pi}/{m}$ \cite{guan:085120}.
At higher temperatures, $T\sim \mu_b$ or $\mu_u$,
the particle and hole excitations are no longer restricted near
{the Fermi surface. Their dispersion} undergoes a crossover from
relativistic (linear in $k$) to non-relativistic ($k^2$).
In particular, this happens in the vicinity of
quantum critical points, where either $\mu_b$ or $\mu_u$ becomes
vanishingly small. {The conformal invariance is now violated due
  to the lack of Lorentz invariance. Consequently,
Eq.~\eqref{conformal} becomes inadequate.}
Nevertheless, we shall show
that the system is still equivalent to two weakly coupled gases
despite the lack of conformal invariance.

At even higher temperature, $T\sim \varepsilon_B$, there is
enough thermal energy to break pairs. Similarly, for $T\sim 2h$, a
leftover fermion can flip its spin to be anti-parallel to the external
field $h$ due to thermal {excitation}.
However, in this paper, we focus on the physically
interesting regime {of low temperatures and strong coupling,}
\be
T\ll \varepsilon_B, 2h\;\;\; \mathrm{and}\;\;\;
\gamma\gg 1\,. \label{reg}
\ee

Our key observation is that the TBA equations can be greatly simplified
in the regime~\eqref{reg} due to the suppression of spin fluctuations.
Mathematically, the infinite set of integral equations for spin rapidities
is analytically tractable for strong attraction and high magnetic field. Their overall effect can
be absorbed into a renormalization of $\mu_u$, which becomes exponentially small
at low $T$ so the feedback effect to the spin rapidity solution can be safely neglected.
This leads to our central result---computing the thermodynamics at finite
temperatures only requires solving the coupled algebraic equations:
\bea
\mu_b&=&\mu+\frac{\varepsilon_B}{2}-\frac{a_1}{4}p_b-a_1p_u,
\label{r1} \\ \mu_u&=&\mu+T\ln
(2\cosh\frac{h}{T})-a_1p_b-\frac{a_1}{4}p_u\mathrm{sech}^2\frac{h}{T},
\label{r2} \\
p_b&=&-\sqrt{\frac{m}{\pi\hbar^2}}T^{3/2}\Li_{\frac{3}{2}}(-e^{2\mu_b/T}),
\label {r3} \\
p_u&=&-\sqrt{\frac{m}{2\pi\hbar^2}}T^{3/2}\Li_{\frac{3}{2}}(-e^{\mu_u/T}).
\label{r4}
\eea
Here, $\Li_s(x)$ is the standard polylogarithm function as a
result of the Fermi-Dirac integral.
As we shall focus on the regime~\eqref{reg},
we can further neglect the exponentially small
contributions from the spin fluctuation. Hence, Eq.~\eqref{r2}
is replaced by
\be \mu_u=\mu+h-a_1p_b. \label{mu_sim} \ee

The thermodynamic potential (per unit length) $\SG$
is then given by $\SG =-p=-(p_b+p_u)$, where $p$ is the pressure. From
$\SG$, it is
straightforward to find the density $n=-\partial \SG /\partial \mu$,
the magnetization
$M=-\partial \SG/\partial h$, the entropy $s=-\partial \SG/\partial T$, and the
compressibility $\kappa=\partial n/\partial \mu$. For these purposes, it is
convenient to first take the corresponding derivative of
Eqs.~\eqref{r1}--\eqref{r4},
use $\partial \Li_s (-e^x)/\partial x=\Li_{s-1}(-e^x)$, and
then solve the resulting
linear equations. For example,
defining $\alpha=-(mT/4\pi\hbar^2)^{1/2}\Li_{{1}/{2}}(-e^{2\mu_b/T})$,
$\beta=-(mT/2\pi\hbar^2)^{1/2}$ $\Li_{{1}/{2}}(-e^{\mu_u/T})$ and
$c=1+a_1\alpha-4a_1^2\alpha\beta$, we have
\be
n=(4\alpha+\beta-7a_1\alpha\beta)/c,\;\;M=\beta(1-3a_1\alpha)/c.
\label{density}
\ee
A detailed analysis of the TBA equations together with the derivation of
Eqs.~\eqref{r1}--\eqref{mu_sim} will be presented elsewhere.

\begin{figure}
\includegraphics[width=0.4\linewidth]{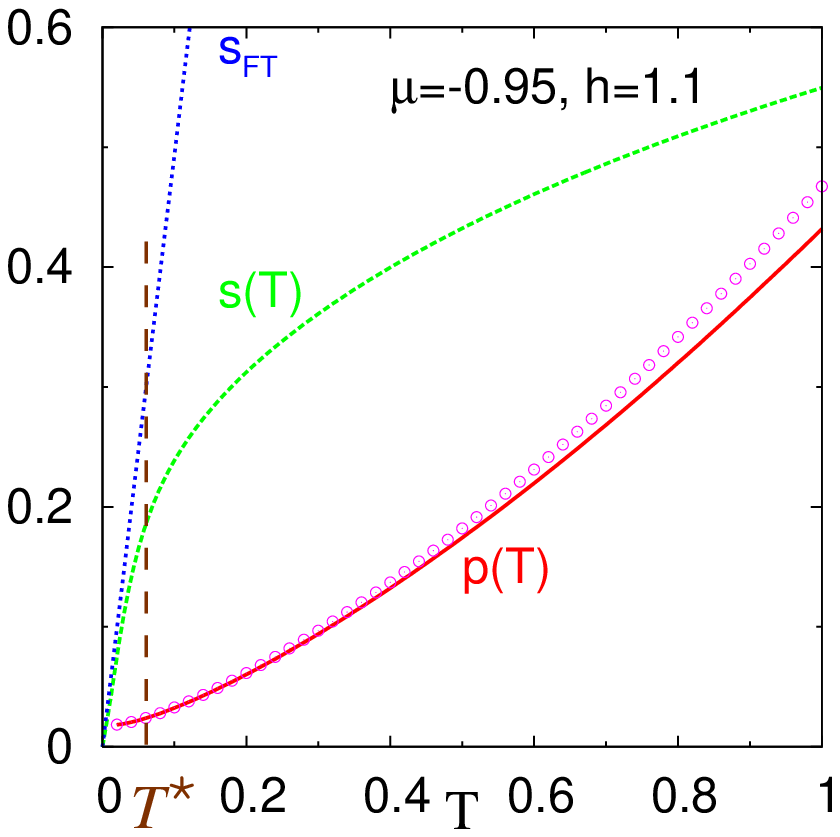}\includegraphics[width=0.5\linewidth]{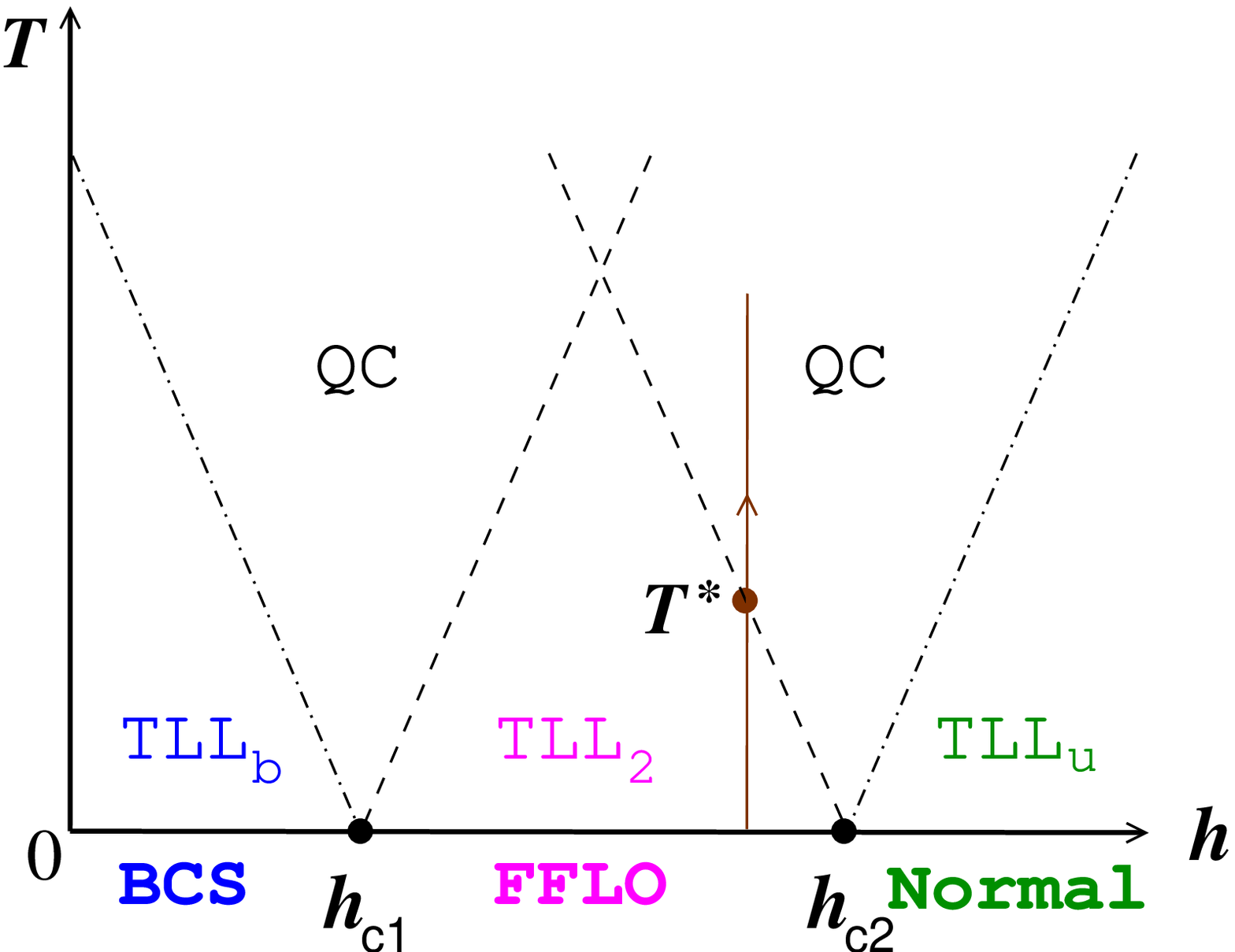}
\caption{ (Color online) {\bf Left panel:} Comparison of the temperature
dependence of the pressure $p$ obtained
by numerically solving the full TBA equations (circles) and
simplified Eqs.~\eqref{r1}--\eqref{r4} (solid line). The entropy $s(T)$ (green
dashed line) is linear in $T$ at low temperatures, agreeing well with
$s_{FT}$ (blue dotted line) as predicted by the low energy effective
field theory. $\mu=-0.95$ and $h=1.1$.
The energy and length units are $\varepsilon_B/2$ and $a_1$,
respectively.  {\bf Right panel:}
The schematic finite-temperature ``phase diagram"
in the regime Eq.~\eqref{reg}.
} \label{fg1}
\end{figure}

For given ($\mu$, $h$, $T$), equations \eqref{r1}-\eqref{mu_sim} can be
solved by simple iteration by treating
$a_1 p_b$ as being small compared
to $\mu_b$ and similarly
$a_1p_u$  as to $\mu_u$.  These equations free us from the
requirement of solving the full TBA equations,
which is a much harder task \cite{kakashvili:041603}.
To test the accuracy of our new approach,
we examined the pressure $p=-\SG$.
In Fig. \ref{fg1}, we compare results
from Eqs.~\eqref{r1}--\eqref{mu_sim}  (solid lines) and that obtained by
numerically solving the full TBA equations (circles).
We observe that
indeed only at high temperature, $T \sim \varepsilon_B/2$, the deviation
from the exact result becomes significant.
Thus the validity of our new formulation is established.

In the low temperature limit $T\ll \mu_{b}, \mu_{u}$
we calculate the free energy using
a Sommerfeld expansion based on Eqs.~\eqref{r1}--\eqref{r2}.
Indeed, we find the thermodynamic
potential 
reproduces the field theory result [Eq.~\eqref{conformal}] and the
corresponding entropy (per unit length)
\be
s_{FT}=\frac{\pi}{3\hbar}(\frac{1}{v_b}+\frac{1}{v_u})T+...\,. \label{efs}
\ee
The entropy $s(T)$ obtained from Eqs.~\eqref{r1}--\eqref{r4} is
shown in Fig. \ref{fg1} together with the low energy effective field
theory result $s_{FT}$.  $s(T)$ is clearly linear in $T$ at low
temperatures and agrees well with $s_{FT}$.  The deviation from the
linear $T$ dependence 
marks a universal crossover around $T^*$ from
relativistic  to non-relativistic dispersion for
particle- and hole-like excitations.
This leads to a minima in the magnetization as well as in
the densities (not shown in Fig. \ref{fg1}) at roughly the same
temperature scale $T^*$.
Analogous crossover phenomena were discussed in
the magnetization curve for gapped spin-1 Heisenberg chains
\cite{maeda:057205}
and experimentally observed in
two-leg spin ladders \cite{kramer2008}.

The gross features of our findings are summarized in the schematic
``phase diagram'' in Fig. \ref{fg1}.  There is no phase transition at
finite temperature but only crossovers between different regimes. At
low temperatures, the BCS, normal, and FFLO phase become the
relativistic TLL of bound pairs (TLL$_b$), unpaired fermions
(TLL$_u$), and two-component TLL (TLL$_2$) respectively.  The TLL
description of the FFLO phase breaks down at crossover temperature
$T^*$ where the dispersion of either bound pairs or leftover fermions
becomes non-relativistic. In particular, in the vicinity of the
quantum critical points ($h_{c1}$ and $h_{c2}$), the system crosses
over to the ``quantum critical (QC) regime''. The other crossover
(dash-dot) lines roughly correspond to the excitation gap for unpaired
fermions or bound pairs.

{\sf Application: two-stage thermometry.}
Now we use these formulas to study the strongly interacting
imbalanced Fermi gas inside a 1D harmonic trap of frequency $\omega$.
According to the local density approximation, the chemical potential
varies in space as $\mu(x)=\mu_0 - \frac{1}{2}m\omega^2x^2$, with
$x=0$ corresponding to the center of the trap, while the magnetic
field $h$ stays constant. Thus, distinct zero temperature phases
are realized at different locations in the trap, giving rise to particular
spatial structures in the in-situ density images.
A central challenge in interpreting the density distribution data
from experiments is how to determine ($T, \mu_0, h$), none of which
seems to be accurately measurable so far.

We use our results to propose a two-stage scheme to accurately determine the
parameters $T$, $\mu_0$, and $h$ from the density profiles. {\it Stage 1:}
estimate the values of these parameters using the density
profiles at the phase boundary and at the edge of the gas
cloud. {\it Stage 2:} using the estimations as guess inputs, obtain ($T,
\mu_0, h$) to high accuracy by a direct three-parameter fit of the
whole density profile using Eq.~\eqref{density}.
Previously, thermal tails at the outer wing
of the density cloud have been used to extract the temperature of
atomic gases in 3D traps based on the observation that the gas is
essentially non-interacting there
\cite{shin2008,ho-zhou-den}.  Here, by contrast, the
outer wing may still interact strongly (e.g., it might be a
Tonks-Girardeau gas). Moreover, our method utilizes densities not only
at the edge of the cloud, but also at the interface. We now
illustrate how to accomplish the goals of Stage 1 for the cases of
low and high total polarization.

\begin{figure}
\includegraphics[width=0.5\linewidth]{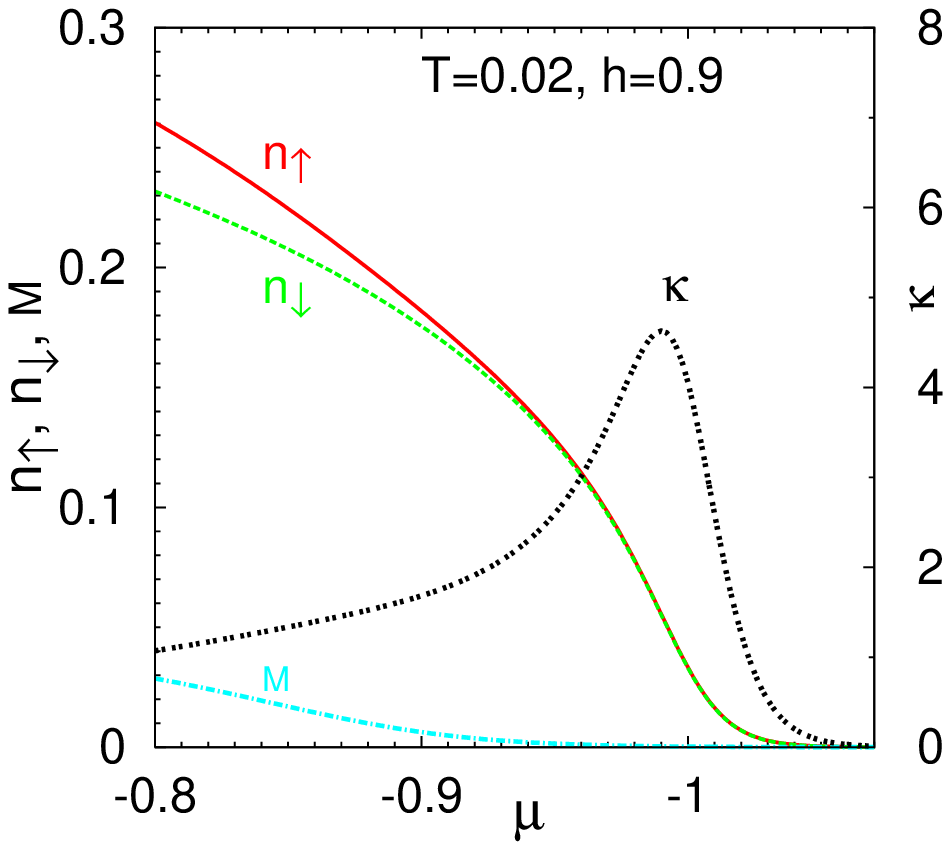}\includegraphics[width=0.5\linewidth]{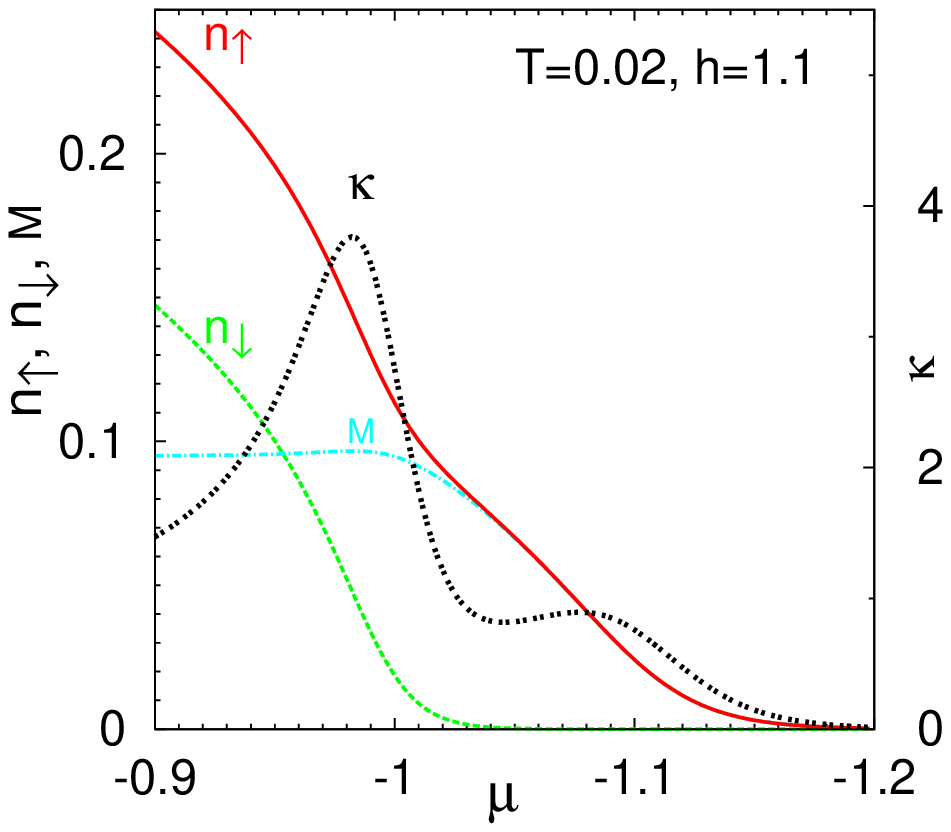}
\caption{ The density $n_\ua$, $n_\da$, the magnetization
$M=n_\ua-n_\da$, and the compressibility $\kappa=\partial n/\partial
\mu$ versus the local chemical potential $\mu$ in a 1D trap. {\bf Left
panel:} low polarization, $h=0.9$.  {\bf Right panel:} high
polarization, $h=1.1$. The temperature $T=0.02$ (units are the same as
Fig. \ref{fg1}). $\kappa$ develops a peak at the thermal tail of the
gas or at the phase boundary. Its peak value depends sensitively on
temperature. }
\label{fb}
\end{figure}

For {\it low polarization}, the trap consists of a partially polarized
(FFLO) core and a fully paired (BCS) wing, which at strong coupling
is a Tonks-Girardeau gas.
Representative finite $T$ density distributions are
shown in the left panel of Fig. \ref{fb}.
At $T=0$, the magnetization $M$ drops to zero
at the FFLO-BCS phase boundary $\mu=-h+n^3\pi^2a_1/24$ \cite{guan:085120},
while $n_\ua$ and $n_\da$ vanish at $\mu=-\varepsilon_B/2$.
Finite temperature leads to thermal tails for all three.
The compressibility $\kappa$ is also shown in Fig. \ref{fb}, which develops
a peak near the edge of the cloud. At $T=0$, $\kappa$ diverges as
$\sim 1/\sqrt{\mu +\varepsilon_B/2}$
as $\mu$ approaches $-\varepsilon_B/2$.
At finite temperature, this divergence is replaced by a peak which is
increasingly shifted away from $\mu=-\varepsilon_B/2$ and becomes less
pronounced as the
temperature is raised. In the BCS region, $p_u$ is essentially zero,
and the equation of density [Eq.~\eqref{density}]
simplifies to $n=4\alpha/(1+a_1\alpha)$. An analytic
formula for $\kappa$ follows with its peak value given by
the simple expression
\be
\kappa_\mathrm{pk}=0.712[\sqrt{\frac{2m}{\hbar^2T}}
  -\frac{1.29}{\epsilon_Ba_1}+ \ldots ].  \label{eq:k_pk}
\ee
For strong interaction and low temperatures, the
second and higher order terms can be neglected. This formula provides a
powerful way to estimate $T$, namely the peak of compressibility
is a sensitive thermometer.
In fact, $T$, $\mu_0$, and $h$ can be determined from
experimentally measured $n(x)$ and $M(x)$, in the following steps.
\begin{inparaenum}[(A)]
\item Obtain the profile of compressibility from the density
  distribution in the trap, $\kappa(x)=-(m\omega^2x)^{-1}\partial
  n/\partial x$.
\item Read off the position, $x_\mathrm{pk}$, and the magnitude
  of the compressibility peak,
  $\kappa_\mathrm{pk}$, from the plot $\kappa(x)$.
\item Calculate $T$ from $\kappa_\mathrm{pk}$, using
  Eq.~\eqref{eq:k_pk}.
\item Estimate the value of the chemical potential $\mu_0$ at
  the center of the trap from $\mu_b(x_\mathrm{pk})=\mu_0 -
  m\omega^2x^2_\mathrm{pk}/2 +\varepsilon_B/2  \simeq 0.554T$.
\item Estimate $h$ using $h=-\mu_0+m\omega^2x_M^2/2$, where
 $x_M$ is read off from the magnetization profile $M(x)$ at the value of
  $M(x_M)=0.605\sqrt{mT/2\pi\hbar^2}$. Note that the thermal tail of
  $M$ is usually very broad, because $\mu_u\ll \mu_b$ is typically
  smaller than $T$ for low polarization.
\end{inparaenum}

For {\it high polarization}, a typical finite $T$ density profile is
shown in the
right panel of Fig.~\ref{fb}.
At zero temperature, the trap consists of a
partially polarized (FFLO) core and fully polarized normal (N) wing,
with $n_\da$ decaying to zero at the FFLO-N phase boundary.  Thermal
fluctuations smear the $T=0$ phase boundary, but the compressibility
develops a pronounced peak at $x=x_{p1}$ due to the rapid decay of
$n_\da$ at low temperatures. In the normal region at the outer wing,
the fully polarized gas is free and we found that
the compressibility has the simple
form $\kappa=-(m/2\pi\hbar^2 T)^{1/2}\Li_{-1/2}(-e^{\mu_u/T})$. In
this region, the thermal tail of $n_\ua$ leads to a second peak in the
compressibility at $x=x_{p2}$. The two-peak structure of the
compressibility is clearly seen in Fig.~\ref{fb}.
The first stage estimation of ($T,\mu_0,h$) in this case involves three steps.
\begin{inparaenum}[(A')]
\item Measure the value of compressibility, $\kappa_{p2}$, at the
  second (outer) peak,
and obtain the
temperature from the relation
\be
\kappa_{p2}=0.126\sqrt{\frac{2m}{\hbar^2 T}}.
\ee
\item Estimate $\mu_0$ from the first (inner) peak location $x_{p1}$, using
$\mu_b(x_{p1})=\mu_0 - m\omega^2x^2_{p1}/2 +\varepsilon_B/2 \simeq 0.554T$.
\item Estimate $h$ from the second peak location
$x=x_{p2}$ using $\mu_u(x_{p2})=\mu_0-m\omega^2x^2_{p2}+h=1.11T$.
\end{inparaenum}

The authors thank Carlos Bolech, Jason Ho, Randy Hulet, and Qi Zhou for helpful
discussions. The work is supported by the DARPA OLE Program and ARO
(EZ and WVL), the Australian Research Council (XWG and MTB), and MEXT
of Japan (MO).

\bibliography{thermo_atoms}

\begin{thebibliography}{22}
\expandafter\ifx\csname natexlab\endcsname\relax\def\natexlab#1{#1}\fi
\expandafter\ifx\csname bibnamefont\endcsname\relax
  \def\bibnamefont#1{#1}\fi
\expandafter\ifx\csname bibfnamefont\endcsname\relax
  \def\bibfnamefont#1{#1}\fi
\expandafter\ifx\csname citenamefont\endcsname\relax
  \def\citenamefont#1{#1}\fi
\expandafter\ifx\csname url\endcsname\relax
  \def\url#1{\texttt{#1}}\fi
\expandafter\ifx\csname urlprefix\endcsname\relax\def\urlprefix{URL }\fi
\providecommand{\bibinfo}[2]{#2}
\providecommand{\eprint}[2][]{\url{#2}}

\bibitem[{\citenamefont{Zwierlein et~al.}(2006)\citenamefont{Zwierlein,
  Schirotzek, Schunck, and Ketterle}}]{Zwierlein06}
\bibinfo{author}{\bibfnamefont{M.~W.} \bibnamefont{Zwierlein}},
  \bibinfo{author}{\bibfnamefont{A.}~\bibnamefont{Schirotzek}},
  \bibinfo{author}{\bibfnamefont{C.~H.} \bibnamefont{Schunck}},
  \bibnamefont{and} \bibinfo{author}{\bibfnamefont{W.}~\bibnamefont{Ketterle}},
  \bibinfo{journal}{Science} \textbf{\bibinfo{volume}{311}},
  \bibinfo{pages}{492} (\bibinfo{year}{2006}).

\bibitem[{\citenamefont{Partridge et~al.}(2006)\citenamefont{Partridge, Li,
  Kamar, Liao, and Hulet}}]{Partridge06}
\bibinfo{author}{\bibfnamefont{G.~B.} \bibnamefont{Partridge}},
  \bibinfo{author}{\bibfnamefont{W.}~\bibnamefont{Li}},
  \bibinfo{author}{\bibfnamefont{R.~I.} \bibnamefont{Kamar}},
  \bibinfo{author}{\bibfnamefont{Y.-A.} \bibnamefont{Liao}}, \bibnamefont{and}
  \bibinfo{author}{\bibfnamefont{R.~G.} \bibnamefont{Hulet}},
  \bibinfo{journal}{Science} \textbf{\bibinfo{volume}{311}},
  \bibinfo{pages}{503} (\bibinfo{year}{2006}).

\bibitem[{\citenamefont{Parish et~al.}(2007{\natexlab{a}})\citenamefont{Parish,
  Marchetti, Lamacraft, and Simons}}]{simons}
\bibinfo{author}{\bibfnamefont{M.~M.} \bibnamefont{Parish}},
  \bibinfo{author}{\bibfnamefont{F.~M.} \bibnamefont{Marchetti}},
  \bibinfo{author}{\bibfnamefont{A.}~\bibnamefont{Lamacraft}},
  \bibnamefont{and} \bibinfo{author}{\bibfnamefont{B.~D.}
  \bibnamefont{Simons}}, \bibinfo{journal}{Nat. Phys.}
  \textbf{\bibinfo{volume}{3}}, \bibinfo{pages}{124}
  (\bibinfo{year}{2007}{\natexlab{a}}).

\bibitem[{\citenamefont{Parish et~al.}(2007{\natexlab{b}})\citenamefont{Parish,
  Baur, Mueller, and Huse}}]{parish:07}
\bibinfo{author}{\bibfnamefont{M.~M.} \bibnamefont{Parish}},
  \bibinfo{author}{\bibfnamefont{S.~K.} \bibnamefont{Baur}},
  \bibinfo{author}{\bibfnamefont{E.~J.} \bibnamefont{Mueller}},
  \bibnamefont{and} \bibinfo{author}{\bibfnamefont{D.~A.} \bibnamefont{Huse}},
  \bibinfo{journal}{Phys. Rev. Lett.} \textbf{\bibinfo{volume}{99}},
  \bibinfo{eid}{250403} (\bibinfo{year}{2007}{\natexlab{b}}).

\bibitem[{\citenamefont{Zhao and Liu}(2008)}]{zhao:08}
\bibinfo{author}{\bibfnamefont{E.}~\bibnamefont{Zhao}} \bibnamefont{and}
  \bibinfo{author}{\bibfnamefont{W.~V.} \bibnamefont{Liu}},
  \bibinfo{journal}{Phys. Rev. A} \textbf{\bibinfo{volume}{78}},
  \bibinfo{eid}{063605} (\bibinfo{year}{2008}).

\bibitem[{\citenamefont{Yang}(2001)}]{KunYang:01}
\bibinfo{author}{\bibfnamefont{K.}~\bibnamefont{Yang}}, \bibinfo{journal}{Phys.
  Rev. B} \textbf{\bibinfo{volume}{63}}, \bibinfo{pages}{140511}
  (\bibinfo{year}{2001}).

\bibitem[{\citenamefont{Moritz et~al.}(2005)}]{PhysRevLett.94.210401}
\bibinfo{author}{\bibfnamefont{H.}~\bibnamefont{Moritz}} \bibnamefont{et~al.},
  \bibinfo{journal}{Phys. Rev. Lett.} \textbf{\bibinfo{volume}{94}},
  \bibinfo{pages}{210401} (\bibinfo{year}{2005}).

\bibitem[{\citenamefont{Hulet}(2009)}]{randy}
\bibinfo{author}{\bibfnamefont{R.~G.} \bibnamefont{Hulet}}
  (\bibinfo{year}{2009}), \bibinfo{note}{private communications}.

\bibitem[{\citenamefont{Takahashi}(1999)}]{Takahashi}
\bibinfo{author}{\bibfnamefont{M.}~\bibnamefont{Takahashi}},
  \emph{\bibinfo{title}{Thermodynamics of One-Dimensional Solvable Models}}
  (\bibinfo{publisher}{Cambridge University Press},
  \bibinfo{address}{Cambridge}, \bibinfo{year}{1999}).

\bibitem[{\citenamefont{Kakashvili and Bolech}(2009)}]{kakashvili:041603}
\bibinfo{author}{\bibfnamefont{P.}~\bibnamefont{Kakashvili}} \bibnamefont{and}
  \bibinfo{author}{\bibfnamefont{C.~J.} \bibnamefont{Bolech}},
  \bibinfo{journal}{Phys. Rev. A} \textbf{\bibinfo{volume}{79}},
  \bibinfo{eid}{041603} (\bibinfo{year}{2009}).

\bibitem[{\citenamefont{Casula et~al.}(2008)\citenamefont{Casula, Ceperley, and
  Mueller}}]{casula:033607}
\bibinfo{author}{\bibfnamefont{M.}~\bibnamefont{Casula}},
  \bibinfo{author}{\bibfnamefont{D.~M.} \bibnamefont{Ceperley}},
  \bibnamefont{and} \bibinfo{author}{\bibfnamefont{E.~J.}
  \bibnamefont{Mueller}}, \bibinfo{journal}{Phys. Rev. A}
  \textbf{\bibinfo{volume}{78}}, \bibinfo{eid}{033607} (\bibinfo{year}{2008}).

\bibitem[{\citenamefont{Gaudin}(1967)}]{Gaudin196755}
\bibinfo{author}{\bibfnamefont{M.}~\bibnamefont{Gaudin}},
  \bibinfo{journal}{Phys. Lett. A} \textbf{\bibinfo{volume}{24}},
  \bibinfo{pages}{55 } (\bibinfo{year}{1967}).

\bibitem[{\citenamefont{Yang}(1967)}]{PhysRevLett.19.1312}
\bibinfo{author}{\bibfnamefont{C.~N.} \bibnamefont{Yang}},
  \bibinfo{journal}{Phys. Rev. Lett.} \textbf{\bibinfo{volume}{19}},
  \bibinfo{pages}{1312} (\bibinfo{year}{1967}).

\bibitem[{\citenamefont{Orso}(2007)}]{orso:070402}
\bibinfo{author}{\bibfnamefont{G.}~\bibnamefont{Orso}}, \bibinfo{journal}{Phys.
  Rev. Lett.} \textbf{\bibinfo{volume}{98}}, \bibinfo{eid}{070402}
  (\bibinfo{year}{2007}).

\bibitem[{\citenamefont{Hu et~al.}(2007)\citenamefont{Hu, Liu, and
  Drummond}}]{hu:070403}
\bibinfo{author}{\bibfnamefont{H.}~\bibnamefont{Hu}},
  \bibinfo{author}{\bibfnamefont{X.-J.} \bibnamefont{Liu}}, \bibnamefont{and}
  \bibinfo{author}{\bibfnamefont{P.~D.} \bibnamefont{Drummond}},
  \bibinfo{journal}{Phys. Rev. Lett.} \textbf{\bibinfo{volume}{98}},
  \bibinfo{eid}{070403} (\bibinfo{year}{2007}).

\bibitem[{\citenamefont{Guan et~al.}(2007)\citenamefont{Guan, Batchelor, Lee,
  and Bortz}}]{guan:085120}
\bibinfo{author}{\bibfnamefont{X.~W.} \bibnamefont{Guan}},
  \bibinfo{author}{\bibfnamefont{M.~T.} \bibnamefont{Batchelor}},
  \bibinfo{author}{\bibfnamefont{C.}~\bibnamefont{Lee}}, \bibnamefont{and}
  \bibinfo{author}{\bibfnamefont{M.}~\bibnamefont{Bortz}},
  \bibinfo{journal}{Phys. Rev. B} \textbf{\bibinfo{volume}{76}},
  \bibinfo{eid}{085120} (\bibinfo{year}{2007}).

\bibitem[{\citenamefont{Bl\"ote et~al.}(1986)\citenamefont{Bl\"ote, Cardy, and
  Nightingale}}]{PhysRevLett.56.742}
\bibinfo{author}{\bibfnamefont{H.~W.~J.} \bibnamefont{Bl\"ote}},
  \bibinfo{author}{\bibfnamefont{J.~L.} \bibnamefont{Cardy}}, \bibnamefont{and}
  \bibinfo{author}{\bibfnamefont{M.~P.} \bibnamefont{Nightingale}},
  \bibinfo{journal}{Phys. Rev. Lett.} \textbf{\bibinfo{volume}{56}},
  \bibinfo{pages}{742} (\bibinfo{year}{1986}).

\bibitem[{\citenamefont{Affleck}(1986)}]{PhysRevLett.56.746}
\bibinfo{author}{\bibfnamefont{I.}~\bibnamefont{Affleck}},
  \bibinfo{journal}{Phys. Rev. Lett.} \textbf{\bibinfo{volume}{56}},
  \bibinfo{pages}{746} (\bibinfo{year}{1986}).

\bibitem[{\citenamefont{Maeda et~al.}(2007)\citenamefont{Maeda, Hotta, and
  Oshikawa}}]{maeda:057205}
\bibinfo{author}{\bibfnamefont{Y.}~\bibnamefont{Maeda}},
  \bibinfo{author}{\bibfnamefont{C.}~\bibnamefont{Hotta}}, \bibnamefont{and}
  \bibinfo{author}{\bibfnamefont{M.}~\bibnamefont{Oshikawa}},
  \bibinfo{journal}{Phys. Rev. Lett.} \textbf{\bibinfo{volume}{99}},
  \bibinfo{eid}{057205} (\bibinfo{year}{2007}).

\bibitem[{\citenamefont{R\"{u}egg et~al.}(2008)}]{kramer2008}
\bibinfo{author}{\bibfnamefont{C.}~\bibnamefont{R\"{u}egg}}
  \bibnamefont{et~al.}, \bibinfo{journal}{Phys. Rev. Lett.}
  \textbf{\bibinfo{volume}{101}}, \bibinfo{eid}{247202} (\bibinfo{year}{2008}).

\bibitem[{\citenamefont{Shin et~al.}(2008)\citenamefont{Shin, Schunck,
  Schirotzek, and Ketterle}}]{shin2008}
\bibinfo{author}{\bibfnamefont{Y.-I.} \bibnamefont{Shin}},
  \bibinfo{author}{\bibfnamefont{C.~H.} \bibnamefont{Schunck}},
  \bibinfo{author}{\bibfnamefont{A.}~\bibnamefont{Schirotzek}},
  \bibnamefont{and} \bibinfo{author}{\bibfnamefont{W.}~\bibnamefont{Ketterle}},
  \bibinfo{journal}{Nature} \textbf{\bibinfo{volume}{451}},
  \bibinfo{pages}{689} (\bibinfo{year}{2008}).

\bibitem[{\citenamefont{Ho and Zhou}(2009)}]{ho-zhou-den}
\bibinfo{author}{\bibfnamefont{T.-L.} \bibnamefont{Ho}} \bibnamefont{and}
  \bibinfo{author}{\bibfnamefont{Q.}~\bibnamefont{Zhou}}
  (\bibinfo{year}{2009}), \bibinfo{note}{arXiv:0901.0018}.

\end{thebibliography}
\bibliographystyle{apsrev}

\end{document}